\documentclass[prl,twocolumn,showpacs,amstext,amsxtra,amssymb,latexsym,bbm]{revtex4}
\usepackage{graphicx,epsf}
\usepackage{bm,bbm}
\usepackage{color}

\newcommand{\K}{{\boldsymbol K}}
\renewcommand{\k}{{\boldsymbol k}}
\newcommand{\G}{{\boldsymbol G}}

\newcommand{\be}{\begin{equation}}
\newcommand{\ee}{\end{equation}}

\renewcommand{\v}{{\boldsymbol v}}
\renewcommand{\a}{{\boldsymbol a}}

\newcommand{\D}{{\boldsymbol D}}

\newcommand{\q}{{\boldsymbol q}}

\newcommand{\R}{{\boldsymbol R}}

\newcommand{\ep}{\epsilon}

\begin{document}


\title{Merging of Dirac points in a two-dimensional crystal}

  \author{G. Montambaux, F. Pi\'echon, J.-N. Fuchs and M.O. Goerbig}
\affiliation{Laboratoire de Physique des Solides, CNRS UMR 8502,
Universit\'e Paris-Sud, 91405- Orsay, France}
\date{October 5, 2009}

\begin{abstract}
We study under which general conditions a pair of Dirac points in
the electronic spectrum of a two-dimensional crystal may merge into
a single one. The merging signals a topological transition between a
semi-metallic phase and a band insulator.
We derive a universal Hamiltonian that describes the physical
properties of the transition, which is controlled by a single
parameter, and analyze the Landau-level spectrum in its vicinity.
This merging may be observed in the organic salt
$\alpha-$(BEDT-TTF)$_2$I$_3$ or in  an optical lattice of cold atoms
simulating deformed graphene.
\end{abstract}


 \maketitle

The recent discovery of graphene has stimulated a great interest in
the physics of the two-dimensional ($2D$) Dirac equation in
condensed matter \cite{review}. The electronic dispersion relation
$\ep(\k)$ vanishes at the contact points between two bands, the
so-called Dirac points $\D$ and $-\D$ (up to an arbitrary reciprocal
lattice vector), around which the electronic spectrum is linear. Due
to the particular hexagonal symmetry of graphene, the two Dirac
points are located at the two  inequivalent corners $\K$ and $\K'$
of the first Brillouin zone (BZ). However, that the Dirac points are
located at high-symmetry points in the BZ is not a necessary
condition, but a rather special case. Indeed, a variation of one of
the three nearest-neighbor hopping parameters make the Dirac points
move away from the corners $\K$ and $\K'$. If the variation is
sufficiently strong, the two Dirac points may even merge into a
single one, which possesses a very particular dispersion relation --
it is linear in one direction while being parabolic in the
orthogonal one. The merging of Dirac points is accompanied by a
topological phase transition from a semi-metallic to an insulating
phase \cite{hase,Zhu,Dietl,Goerbig2008,CastroNeto2,Guinea,Volovik}.

Other physical systems, different from graphene and its particular lattice structure,
exist where Dirac points describe the low-energy properties.
Recent papers have shown that a similar spectrum  may arise in an
organic conductor, the $\alpha-$(BEDT-TTF)$_2$I$_3$ salt  under
pressure \cite{Katayama2006,Kobayashi2007}.
Furthermore, it has been shown that it is possible to
observe massless Dirac fermions with cold atoms in
optical lattices \cite{Zhu,Zhao,Hou2009},   where the motion of the
Dirac points may be induced by changing the intensity of the laser
fields \cite{Zhu}.

In this Letter, we study in a more general manner the motion of
Dirac points within a two-band model that respects time-reversal and
inversion symmetry without being restricted to a particular lattice
geometry. We investigate the general conditions for the merging of
Dirac points into a single one $\D_0$, under variation of the
nearest-neighbor hopping parameters. It is shown that the merging
points may only appear in four special points of the BZ, all of
which are given by half of a reciprocal lattice vector $\D_0=\G/2$.
Furthermore we derive a single effective Hamiltonian that describes
the low-energy properties of the system in the vicinity of the
topological phase transition which accompanies the Dirac-point
merging. The effective Hamiltonian allows us to study the continuous
variation of the Landau-level spectrum from $\propto  \sqrt{Bn}$ in
the semi-metallic to $\propto  B(n+1/2)$ in the insulating phase,
while passing the merging point with an unusual $[B(n+1/2)]^{2/3}$
dependence \cite{Dietl}.

 We consider a  two-band  Hamiltonian for a $2D$
crystal with two atoms per unit cell. Quite generally,
 neglecting for the moment the diagonal terms the effect of which
is discussed at the end of this Letter, the Hamiltonian ${\cal
H}(\k)$ reads:

\be {\cal H}(\k)= \left(%
\begin{array}{cc}
0  & f(\k) \\
  f^*(\k) &  0 \\
\end{array}%
\right) \ . \label{H} \ee The off-diagonal coupling is written as

\be f(\k)= \sum_{m,n} t_{mn} e^{- i \k \cdot \R_{mn}} \ ,
\label{fofk} \ee
where the $t_{mn}$'s are real, a consequence of time-reversal
symmetry ${\cal H}(\k)={\cal H}^*(-\k)$, and   $\R_{mn}= m \a_1 + n
\a_2$ are vectors of the underlying Bravais lattice.

If the  energy dispersion $\ep(\k)=\pm |f(\k)|$ possesses Dirac
points $\D$, they are necessarily located at zero energy, $f(\D)=0$.
From the general expression (\ref{fofk}), it is obvious that these
points $\D$ come in by pairs: as a consequence of time-reversal
symmetry, one has $f(\k)=f^*(-\k)$, and thus if $\D$ is solution of
$f(\k)=0$, so is $-\D$. Quite generally, the position $\D$ can be
anywhere in the BZ and move upon variation  of the band parameters
$t_{mn}$.
Writing $\k=\pm \D+ \q$, the function $f(\k)$ is then linearly
expanded around $\pm \D$ as
\begin{eqnarray} f(\pm \D+\q)&=& - i \q \cdot (\sum_{mn} t_{mn} \R_{mn} \cos \D \cdot \R_{mn})
\nonumber \\ &\pm & \q \cdot (\sum_{mn} t_{mn} \R_{mn} \sin \D \cdot
\R_{mn}) \ ,
 \label{fofq}
\end{eqnarray}
which has the form $\q \cdot (\pm \v_1 - i \v_2)$, and the
linearized Hamiltonian reads: $ {\cal H}_{\pm \D}= \pm \v_1 \cdot \q
\sigma^x + \v_2 \cdot \q \sigma^y$
 in terms of the   Pauli matrices $\sigma^x$ and $\sigma^y$.

Now we consider the situation where, upon variation of the band
parameters, the two Dirac points may approach each other and merge
into a single point $\D_0$. This happens when $\D=-\D$   modulo a
reciprocal lattice vector $\G=p \a^*_1 + q \a_2^*$, where  $\a^*_1$
and  $\a_2^*$ span the reciprocal lattice. Therefore, the location
of this merging point is simply $\D_0= \G/2$. There are then four
possible inequivalent points the coordinates of which are $\D_0= (
p\a^*_1 + q \a_2^*)/2$, with $(p,q)$ = $(0,0)$, $(1,0)$, $(0,1)$,
and $(1,1)$.
 The condition
$ f(\D_0)= \sum_{mn} (-1)^{\beta_{mn}} t_{mn} =0$, where
$\beta_{mn}= p m + q n$, defines a manifold in the space of band
parameters. As we  discuss below, this manifold separates a
semi-metallic phase with two Dirac cones and a band insulator.

Notice that in the vicinity of the $\D_0$ point, $f$ is {\it purely
imaginary} ($\v^0_1=0$), since $\sin (\G \cdot \R_{mn}/2)=0$.
Consequently, to lowest order, the linearized Hamiltonian  reduces
to ${\cal H}= \q
 \cdot \v_2^0  \sigma^y$, where  $\v^0_2=\sum_{mn} (-1)^{\beta_{mn}} t_{mn} \R_{mn} $.
We choose the local  reference system such that $\v^0_2 \equiv c \,
\hat y$ defines the $y$-direction. In order to account for the
dispersion in the local $x$-direction,  we have to expand
$f(\D_0+\q)$ to second order in $\q$:
 \be f(\D_0+\q)= - i \q \cdot \v^0_2 -{1 \over 2} \sum_{mn} (-1)^{\beta_{mn}}
  t_{mn} ( \q \cdot
 \R_{mn})^2 \ .  \label{fofq0} \ee
Keeping the quadratic term in $q_x$, the new Hamiltonian
 may be written as
  ${\cal H}_0(\q)={q_x^2  \over 2 m^*} \sigma^x + c q_y \sigma^y$
where the mass $m^*$ is defined by
 \be {1 \over m^*}=  \sum_{mn} (-1)^{{\beta_{mn}}+1} t_{mn}
 R^2_{mn,x} \ ,  \ee
where  $R_{mn,x}$ is the component of $\R_{mn}$ along   the {\it
local} $x$-axis (perpendicular to $\v_2^0$).  The terms of order
$q_y^2$ and $q_x q_y$ are neglected at low energy.
  The diagonalization of ${\cal H}_0(\q)$ is
straightforward and the energy spectrum $\ep = \sqrt{({q_x^2 / 2
m^*})^2+ c^2 q_y^2}$ has a new structure: it is linear in one
direction and quadratic in the other.   From the linear-quadratic
spectrum which defines a velocity $c$ and a mass $m^*$, one may
identify a characteristic energy :

\be m^* c^2= {[\sum_{mn} (-1)^{\beta_{mn}} t_{mn} \R_{mn} ]^2 \over
\sum_{mn} (-1)^{{\beta_{mn}}+1} t_{mn} R^2_{mn,x} } \ . \ee

\begin{figure}[!h]
\centerline{ \epsfxsize 7cm  \epsffile{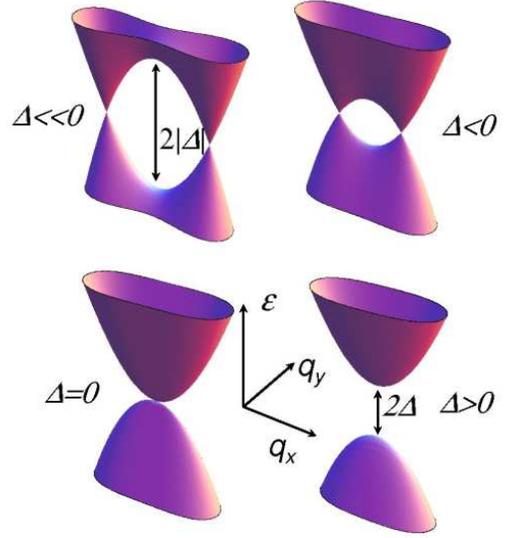}} \caption{\it
Evolution of the spectrum when the quantity $\Delta$ is varied and
changes in sign  at the topological transition (arbitrary units).
The low-energy spectrum  stays linear in the $q_y$ direction.}
\label{fig.bicones}
\end{figure}

Up to now, we have discussed the merging of the two Dirac points
from a ``dynamical'' point of view,  following their motion in the
BZ when varying the band parameters until $\D_0$ is reached. We now
consider the low energy Hamiltonian around $\D_0$ even before the
two Dirac points coincide. In the neighborhood of the transition
when $f(\D_0)=0$, there is a finite gap $2 |\Delta|$ {\it at} the
$\D_0$ point (see Fig. \ref{fig.bicones}), where the quantity
\be \Delta= \sum_{mn} (-1)^{\beta_{mn}} t_{mn} \label{gap} \ee
changes its sign at the transition. This parameter $\Delta$
therefore drives the transition. In the vicinity of $\D_0$, the
Hamiltonian becomes  ${\cal H}(\q)={\cal H}_0(\q)+ \Delta \sigma^x$,
or explicitly~:

\be {\cal H}(\q)= 
\left(
  \begin{array}{cc}
    0 & \Delta+ {q_x^2 \over 2 m^*} -  i c q_y  \\
 \Delta + {q_x^2 \over 2 m^*} +  i c q_y & 0 \\
  \end{array}
\right) \label{newH} \ee
with the spectrum $\ep= \pm \sqrt{(\Delta + q_x^2/2 m^*)^2+ c^2
q_y^2}$. The Hamiltonian ({\ref{newH}) has a remarkable structure
and describes properly the vicinity of the topological transition,
as shown on Fig. \ref{fig.bicones}. When $ m^* \Delta$ is negative
(we  choose $m^* >0$ without loss of generality), the spectrum
exhibits the two Dirac cones and a saddle point in $\D_0$ (at half
distance between the two Dirac  points). Increasing $\Delta$ from
negative to positive values, the saddle point evolves into the
hybrid point  at the transition   ($\Delta=0$) before
 a gap $2 \Delta >0$ opens.
Due to the linear spectrum near the Dirac points, the density of
states in the semi-metallic phase varies as $|\ep|$ at low energy
and exhibits a logarithmic divergence $\ln( ||\ep| - |\Delta||)$ due
to the saddle point. At the transition, it varies as $\sqrt{|\ep|}$
and then a gap opens for $\Delta >0$ \cite{Dietl}.

The topological character of the transition is displayed by  the
cancelation of the Berry phase at the merging of the  two Dirac
points. The spinorial structure of the wave function
 leads to a  Berry phase   ${1 \over 2} \oint
 \nabla \theta_\q \cdot d \q $, where   $\theta_\q =\arctan {\mbox{Im} f(\q) \over
\mbox{Re} f(\q)}$}. Near each Dirac point,  $\theta_\q =\arctan {q_y
\over q_x}$, whereas  $\theta_\q = \arctan {2 m^* c q_y \over
q_x^2}$ near the hybrid $\D_0$ point at the transition. Therefore,
the Berry phases $\pm \pi$ around each Dirac point annihilate when
they merge into $\D_0$ \cite{Dietl}.

\begin{figure}[!h]
\centerline{ \epsfxsize 4cm \epsffile{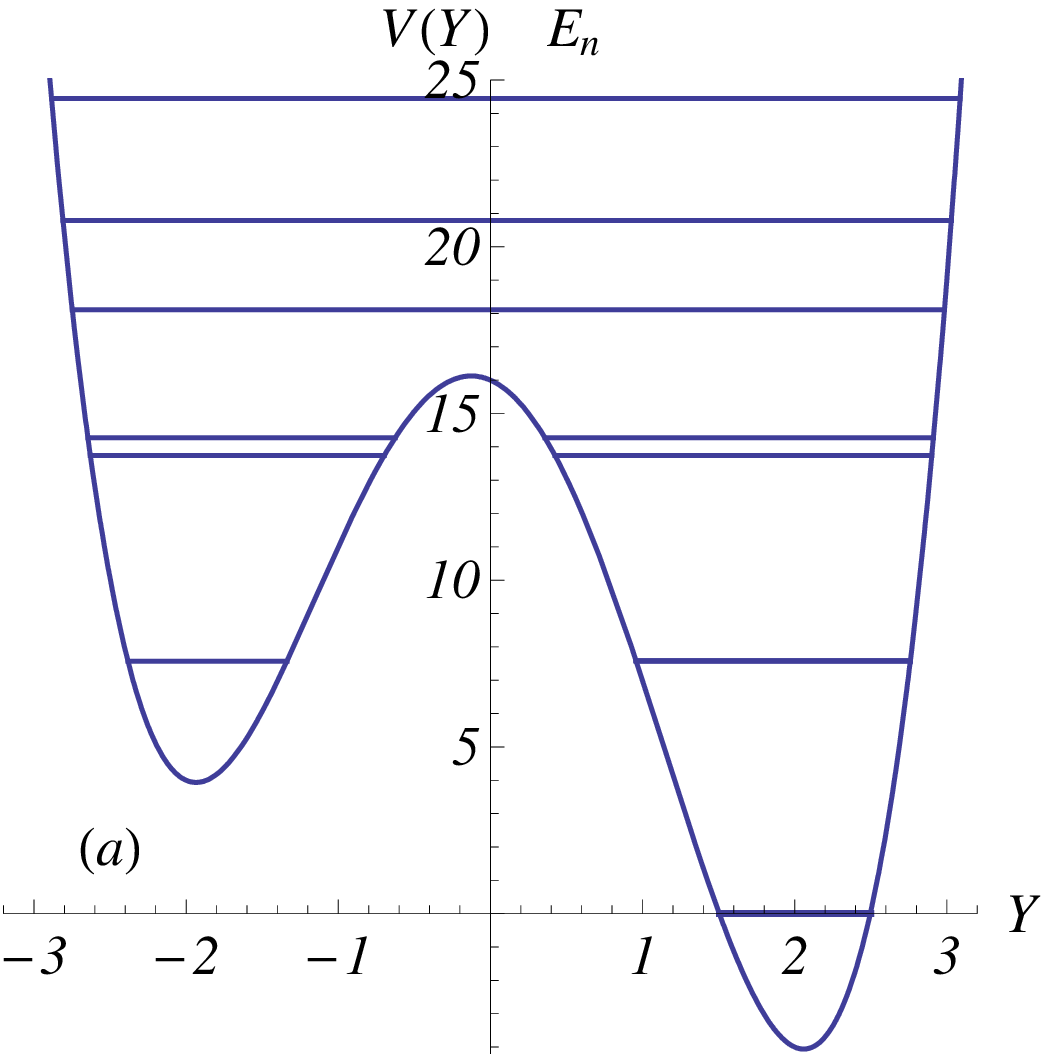}  \epsfxsize 4cm
\epsffile{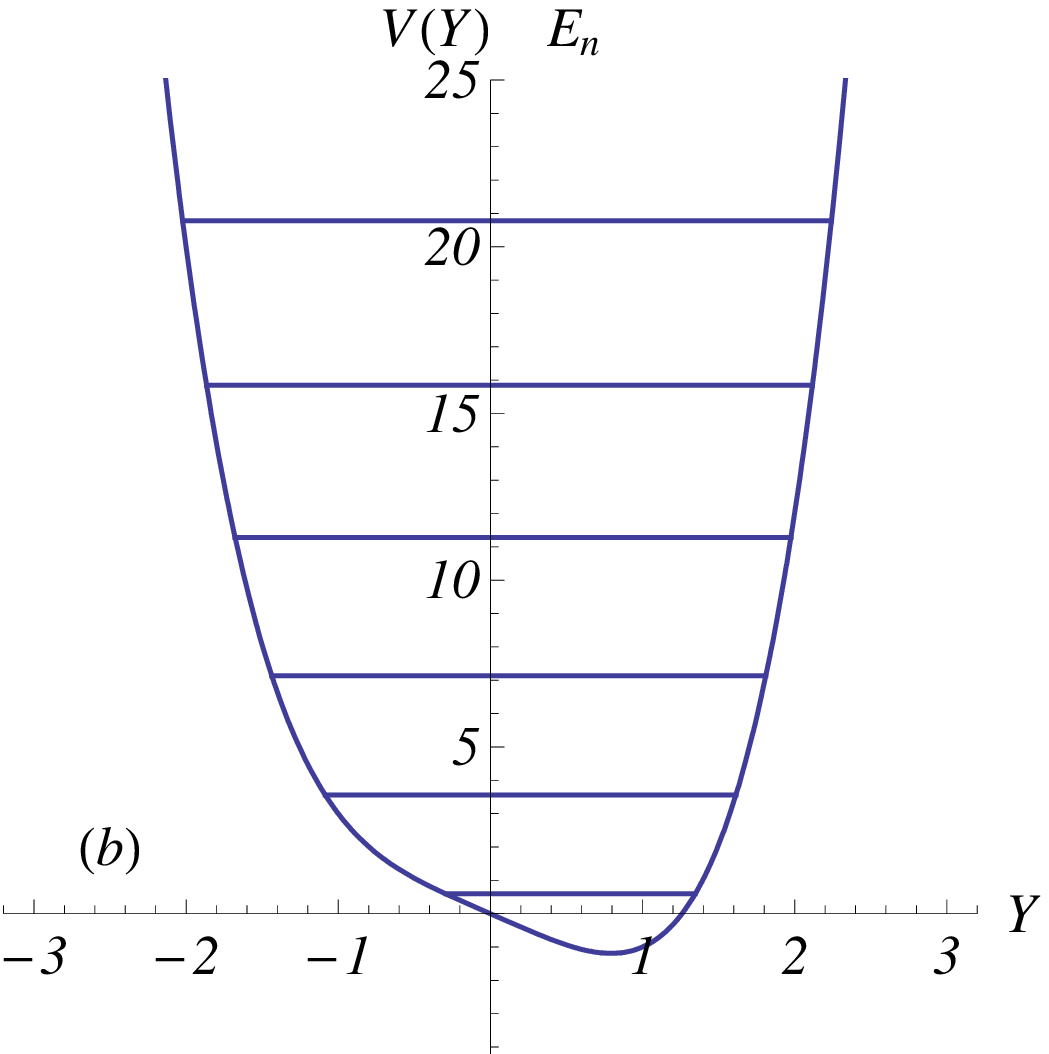}} \caption{\it Potential profile $V(Y)=(\delta +
Y^2)^2 - 2 Y$ and effective energy levels $E_{n}$
 for
$\delta=-4$ (a) and $\delta=0$ (b). } \label{paraboles}
\end{figure}

We now turn to the evolution of the spectrum in a perpendicular
magnetic field $B$. After the substitution $q_x \rightarrow q_x - e
B y$ in the appropriate gauge  and the introduction of the
dimensionless gap $\delta= \Delta/(m^* c^2 \omega_c^2 /2)^{1/3}$,
in terms of the cyclotron frequency $\omega_c=eB/m^*$, the
eigenvalues  are $\ep_n = \pm ({\Delta / \delta})
\sqrt{E_n(\delta)}$ where the $E_n$ are solutions of the {\it
effective} Schr\"odinger equation
\be E_n(\delta) \psi = \big[ P^2 + (\delta + Y^2)^2 - 2 Y \big] \
\psi \equiv {\cal H}_{eff} \psi \ee
 with $[Y,P]=i$. The effective Hamiltonian is therefore of Schr\"odinger type with  a
 double-well potential when $\delta <0$, which becomes the quartic
 potential $Y^4-2 Y$ at the transition and then acquires a gap for
 $\delta >0$, with a parabolic dispersion at low energy (see  Fig.
 \ref{paraboles}). For large negative $\delta$, one recovers two
 independent parabolic wells with an energy shift $\pm 2 \sqrt{\delta}$
equal to  half the cyclotron energy. Therefore, as seen in  Fig.
\ref{paraboles}(a), the lowest level has zero energy, and the first
levels are degenerate : one recovers the physics of independent
Dirac cones in a magnetic field, and the effective energy levels are
given by $E_n= 4 n\sqrt{|\delta|}$.

\begin{figure}[!h]
\centerline{ \epsfxsize 7cm \epsffile{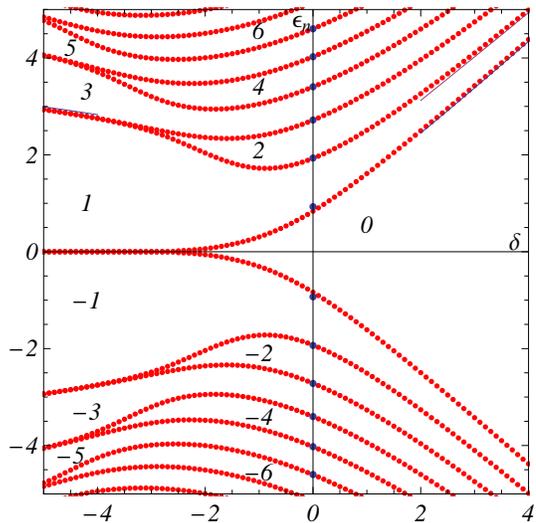}} \caption{\it Energy
levels   $\ep_n(\delta)/(m^* c^2  \omega^2 /2)^{1/3}$ as a function
of the dimensionless parameter $\delta \propto \Delta/B^{2/3}$. The
dots on the $\delta=0$ axis indicate the semi-classical levels of
the quartic Hamiltonian \cite{Dietl}.} \label{fig.landau}
\end{figure}

 The complete  Landau levels spectrum  $\ep_n(\delta)$ is shown  in Fig.
 \ref{fig.landau}. The value of the Hall integer is indicated in the gaps between Landau levels.
 For negative $\delta$, one recovers the
 spectrum of the Dirac cones, with odd values of the Hall integers  -- the absence of even values reflects
 the twofold valley degeneracy of the Landau levels and the presence of a zero-energy Landau level. When $- \delta$
 vanishes, approaching the transition, the level degeneracy is
 lifted, and gaps with even Hall integer  open. A simple WKB analysis of the lowest level
 shows that it splits as  $\ep_0 \propto  \pm e^{-\# |\delta|^{3/2}} \simeq
 \pm
 e^{-\#|\Delta|^{3/2}/B}$.
The energy levels scale as   $\ep_n \propto  B^{2/3}
\sqrt{E_n(\Delta/B^{2/3})}$, with the following limits

\begin{eqnarray}
m^* \Delta <0 \quad , \quad \mbox{semi-metal}  &\rightarrow& \ep_n
\propto \pm \sqrt{n B}
  \\
\Delta =0 \quad , \quad  \mbox{transition}   &\rightarrow& \ep_n
\propto \pm [(n+1/2)B]^{2/3}
  \nonumber  \\
m^* \Delta > 0 \quad , \quad  \mbox{insulator}  &\rightarrow& \ep_n
=\pm [ \Delta + \# (n +1/2) B]
   \nonumber 
\end{eqnarray}
 Note the shift $n \rightarrow n+1/2$, a consequence of the
$\pm \pi$ Berry phases.

 We now consider two specific situations in which the merging of Dirac
points may be observed.
 The first example is a variation of the standard graphene
tight-binding model, where the three hopping integrals between
nearest carbon atoms are  assumed to be different:

\be f(\k)= t_{00} + t_{10} e^{-i \k \cdot \a_1} + t_{01} e^{-i \k
\cdot \a_2} \ . \ee
 A merging at  $\D_0= (p \a_1^* + q \a_2^*)/2$ is possible
if

\be  t_{00} + (-1)^p t_{10}  + (-1)^q t_{01}=0 \ .
\label{D0graphene} \ee

Choosing $t_{00}=t' >0$ and $t_{10}=t_{01}=t >0$, Eq.
(\ref{D0graphene}) has a solution ($t'=2 t$) for $p=q=1$, at
$\D_0=(\a_1^*+\a_2^*)/2$, that is at the $M$ point located at the
edge center of the BZ \cite{Dietl}.   Even if the hopping integrals
may be modified in graphene under uniaxial stress
\cite{Goerbig2008,CastroNeto2}, it seems impossible to reach
physically the merging condition. An alternative for the observation
of Dirac points has been proposed with
 cold atoms in a honeycomb optical lattice. The latter can be
 realized with laser beams, and by changing the amplitude of the
 beams, it is possible to vary the band parameters and to reach a
 situation where the Dirac points merge \cite{Zhu}.

The organic conductor $\alpha-$(BEDT-TTF)$_2$I$_3$ is also a good
candidate for the observation of  merging Dirac points. In order to
study the low-energy spectrum (close to half-filling), the original
description with four molecules per unit cell can be reduced   to a
two-band model in a tetragonal lattice, with the following
dispersion relation \cite{Hotta,Katayama2006,Kobayashi2007}:

\be f(\k)= t_{00} + t_{10} e^{i \k \cdot \a_1} + t_{01} e^{i \k
\cdot \a_2} + t_{11} e^{i \k \cdot (\a_1+ \a_2)} \ .\label{japonais}
\ee
In this case, the generic spectrum   exhibits two Dirac cones the
positions of which are given by \cite{Goerbig2008}

\begin{eqnarray}
\tan^2 {\D \cdot \a_1 \over 2} = {(t_{00}+t_{01})^2 - (t_{11}+
t_{10})^2 \over
(t_{11}- t_{10})^2 - (t_{00} - t_{01})^2} \nonumber \\
\tan^2 {\D \cdot \a_2 \over 2} = {(t_{00}+t_{10})^2 - (t_{11}+
t_{01})^2 \over (t_{11}- t_{01})^2 - (t_{00} - t_{10})^2} \ .
\label{coordinates}
\end{eqnarray}

\begin{figure}[!h]
\centerline{ \epsfxsize 4cm \epsffile{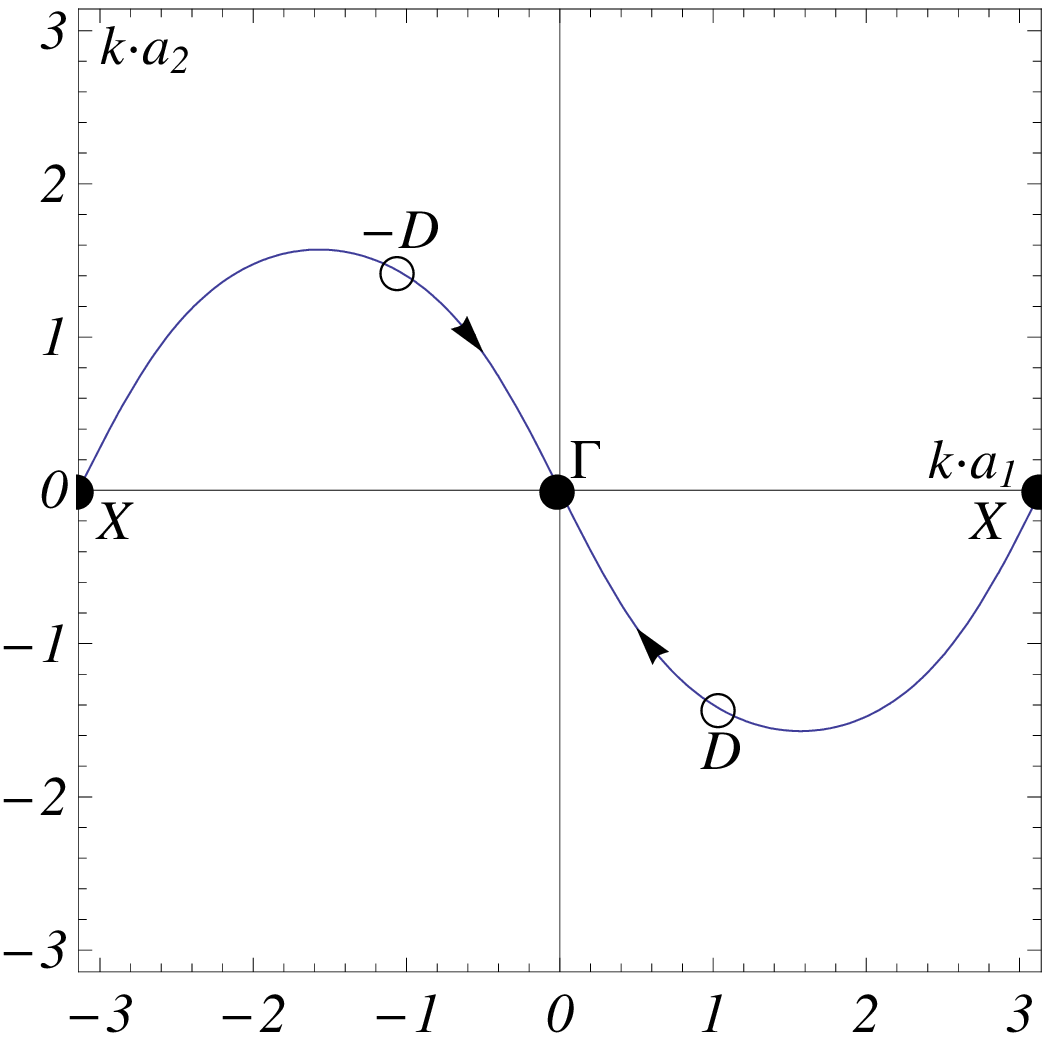} \epsfxsize 4cm
\epsffile{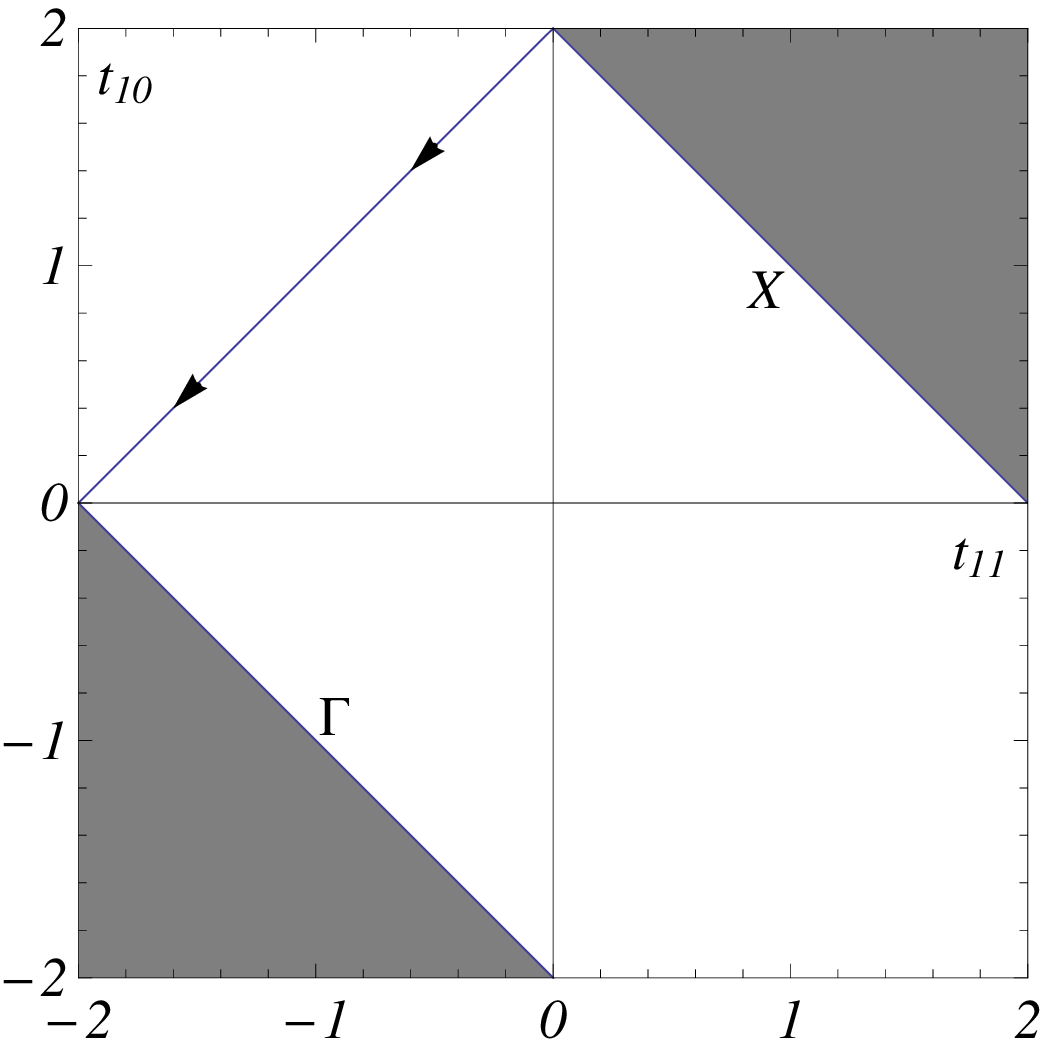}} \centerline{ \epsfxsize 8cm
\epsffile{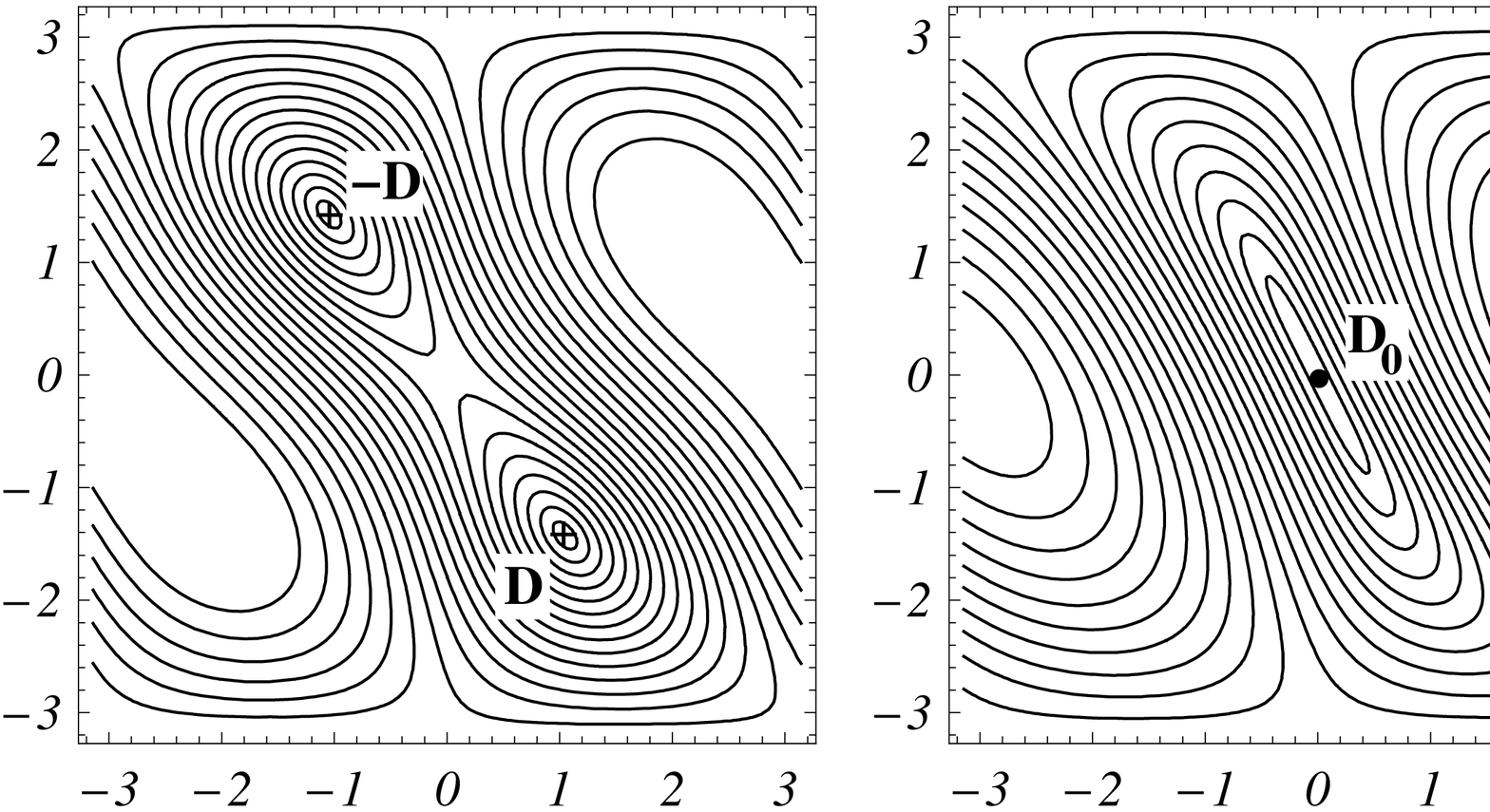} } \caption{\it Top figures: left) motion of the
two Dirac points $\D$ and $-\D$ for the
 case $t_{00}=t_{01}=1$, $t_{10}=2 + t_{11}$ under variation of
$t_{11}$; right) phase diagram. The $\Gamma$ and $X$ lines separate
the semi-metallic phase from the band insulator (grey). The Dirac
points move from  $X=(0,1)$  to the $\Gamma=(0,0)$ point when
$t_{11}$ varies from $0$ to $-2$. Bottom figures show the iso-energy
curves for $t_{11}=-1.5$ (left) and $-2$ (right).
}\label{mouvementD}
\end{figure}

Upon variation of the band parameters, the two Dirac points may
merge when
\be
  t_{00} + (-1)^p t_{10} + (-1)^q t_{01}   + (-1)^{p+q} t_{11}=0  \
  .
\label{carre} \ee
Katayama {\it et al.} have considered the situation (in our
notations) where $t_{00}=t_{01}$
 \cite{Katayama2006} and shown the possibility of a transition from
a massless ``Dirac" phase to a gapped phase at  a hydrostatic
pressure
 $\sim 40$
kbar \cite{Kobayashi2007}. In Fig. \ref{mouvementD}, we show the
evolution of the Dirac points in the  BZ (as in Ref.
\cite{Katayama2006}), and more important, the evolution of the
spectrum for a  particular variation of the band parameters. The two
Dirac points merge at the  BZ points $\Gamma$ and $X$ for special
values of the band parameters.

\begin{figure}[!h]
\centerline{ \epsfxsize 5cm \epsffile{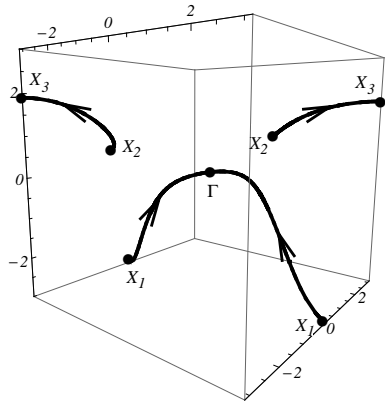}}  \caption{\it
Motion of the Dirac points for model  (\ref{japonais}), with $t_{00}
= -1$, $t_{10} = -0.5$, $t_{01} = 1.4$, while varying $-3 < t_{11}<
3$ (vertical axis).} \label{spaghetti}
\end{figure}

The scenario can be even richer. One may imagine a
 situation where,  when  varying a
band parameter, the Dirac points disappear and then reappear at a
different $\D_0$  point of the BZ. In Fig. \ref{spaghetti}, the
Dirac points move from $X_1$ to $\Gamma$, where a gap opens. For
further variation of the band parameter, the gap persists until a
new pair of Dirac points appears at a different position $X_2$ in
the BZ, and disappears again at the fourth special point $X_3$.

 We finally consider the effect of non-zero  diagonal terms in
the Hamiltonian (\ref{H}). When there is inversion symmetry, one has
${\cal H}_{22}(\k)={\cal H}_{11}(-\k)$. Moreover, time-reversal
symmetry implies that these diagonal matrix elements are symmetric
functions in $\k$, and that their expansion near the hybrid point
$\D_0=\G/2$, has {\it no linear term}. Therefore all considerations
discussed above remain valid, although the Dirac and hybrid points
are no longer necessarily at zero energy.

 In conclusion, we have studied under which general conditions the
merging of Dirac points may occur, marking the transition between a
semi-metal and a band insulator. We have fully described the
vicinity of the transition by means of an effective $2 \times 2$
Hamiltonian. Although it has been constructed to describe the low
energy spectrum near $\D_0$, this Hamiltonian is appropriate to
describe {\it both valleys} around the $\D$ and $-\D$ points
avoiding the use of  a $4 \times 4$ effective Hamiltonian as it is
usually done. It may even provide an effective description of
 graphene, which could be useful, e.g., in
accounting for intervalley scattering in a disordered system.

We recently learned of a related work which  proposed the existence of hybrid
points in the absence of time reversal symmetry in $VO_2/TiO_2$
heterostructures \cite{Pardo2}. We also became aware of a recent independent work on the anisotropic honeycomb lattice
in a magnetic field, which has some overlap with ours \cite{Esaki}. Finally we wish to mention a recent paper on mimicking graphene physics with ultracold fermions in an optical lattice \cite{Lee}.

{\it Acknowledgments - } We acknowledge useful discussions with  S.
Katayama, A. Kobayashi and T. Nishine.

\end{document}